\documentstyle[12pt]{article}

\large
\newcommand{\be}{\begin{eqnarray}}
\newcommand{\ee}{\end{eqnarray}}

\newcommand\noi {\noindent}

\begin{document}
\setlength{\baselineskip}{21pt}
\pagestyle{empty}
\vfill
\eject
\begin{flushright}
SUNY-NTG-93/42
\end{flushright}

\vskip 2.0cm
\centerline{\bf Hot QCD}

\vskip 2.0 cm
\centerline{I. Zahed}
\vskip 1cm
\vskip .5cm
\centerline{Department of Physics}
\centerline{SUNY, Stony Brook, New York 11794}
\vskip 2cm

\centerline{\bf Abstract}

I discuss a comprehensive approach to the spacelike physics in
high temperature QCD in three dimensions. The
approach makes use of dimensional reduction. I suggest
that this approach is useful for high temperature QCD
in four dimensions.

\vfill
\noindent
\begin{flushleft}
SUNY-NTG-93/42\\
October  1993
\end{flushleft}
\eject
\pagestyle{plain}

{\bf 1. Introduction}
\vskip 3mm

The physics of  hot and dense
hadronic matter has received a lot of attention recently. Due to asymptotic
freedom, one expects that at sufficiently high temperatures and/or densities,
hadronic matter in thermal and/or chemical equilibrium will behave as a weakly
interacting system of quarks and gluons (quark-gluon plasma).
As a result, the bulk thermodynamical quantities such as the energy, the
pressure, the entropy, ... should display black-body behaviour. The plasma
should screen for space-like momenta and display collective behaviour
(plasma waves, hydrodynamical waves, ...) for time-like momenta.

The low temperature phase of QCD is dominated by the physics of pions and
the general lore of the spontaneous breaking of chiral symmetry. The high
temperature phase is not. To what extent one expects a phase transition
because of the qualitative change in the ground state properties is not
clear. Lattice simulation results are not yet definitive on this issue.
In so far, they seem to suggest a first order phase transition for four
flavours, and a second order phase transition for two flavours. The real
world with three flavours seems to lie somewhere in between.
For pure Yang-Mills the transition is believed to be first order.

In these notes I will not attempt to address the issue of the phase transition
in QCD. Rather, I will focus on the some of the aspects of the lattice results
at high temperature and try to put forward a general framework for their
global understanding. In section 1, I review
some of the lattice results at high temperature. In section 2, I outline a
general strategy for understanding these results based on dimensional
reduction. In section 3, this strategy is applied to three dimensional QCD
at high temperature. My conclusions are summarized in section 4.

\vskip  1cm
{\bf 2. Lattice Results}
\vskip 3mm

In the vacuum, chiral symmetry is believed to be spontaneously broken.
What this means is that vacuum fluctuations allow for
mixing between left and right handed fermions in the massless case.
In the chiral limit, the order parameter is $<\overline{q}q>$, its
behaviour versus temperature is shown in Fig. 1 for four flavours
[1]. For temperatures of the order of 150 MeV a substantial
decrease in $<\overline{q}q>$ is noted.

In the vacuum, color is expected to be confined at zero temperature.
The order parameter for confinement is the Polyakov loop
$<P>= <{\rm {Tr exp{\bf P}}} \int igA_4 dx_4>$, where the average is a thermal
average, $A_4$
is the imaginary part of the gauge field and the trace is over one period
$\beta =1/T$ of the imaginary time. ${\rm ln}|<P>|$ is usually assumed to
be the free energy of an infinitely heavy quark. A sharp rise in the
expectation value of the Polyakov loop is usually seen following the
drop in the chiral condensate [2].

Lattice measurements of the thermodynamical quantities in QCD show a
rapid cross over in these quantities at about 150 MeV. Fig.2  shows the
behaviour of the energy density and pressure versus $T/T_c$, normalized to the
black-body result [3]. In the regime $T_c < T < (2-3) T_c$, there are
substantial
deviations from the free massless gas limit,
$\Theta^{00} = {\cal E} -3{\cal P}\neq 0$.  The fermionic susceptibilities
(singlet and isotriplet) show also a rapid variation in the same temperature
range as shown in Fig.3 [4]. At low temperature the susceptibilities
are exponentially suppressed by the mass of the hadronic modes.

The Polyakov-anti-Polyakov (connected) correlation function
$<P(x)\,P^+(0)>$ has been measured both at low and high temperature in the
pure Yang-Mills theory. At low temperature, the correlation function displays
an area law behaviour. Recent simulations at high temperature indicate a
screening behaviour [5]. The screening range is estimated to be of the order of
the inverse screening mass $1/gT$. The time-like Wilson loop shows an
area law behaviour at low temperature, and a perimeter law at high
temperature. This is to be contrasted with the space-like Wilson loop
which displays an area law behaviour both at low and high temperature.
The space-like string tension at high temperature has been found to scale
with $T^2$ [5].

Lattice measurements of static hadronic correlation functions show that the
screening lengths asymptote $2\pi T$ (mesons) and $3\pi T$ (baryons) as shown
in Fig.4 [6]. The exception being the pion and the sigma (scalar-isoscalar).
The pattern displayed by the measured screening lengths suggests that the
thermal state preserves chiral symmetry. The pion-sigma channel, however,
suggests the presence of still large fluctuations possibly related with a
chiral phase transition. The lattice measurements of the transverse
correlations in the hadronic channels are shown in Fig.5 , for the pion
and the rho [7]. Almost no change is detected from the low to the high
temperature regime. The correlations are expected to be absent in the free
gas limit.

Recent lattice measurements of the fermionic distribution around the Polyakov
line (heavy quark) are shown in Fig.6 [8]. At low temperature the fermionic
distribution is localized around the heavy source in a range of the order
of $1/\Lambda$ (QCD scale), and sums up to $-1$. At high temperature, the
distribution is considerably smeared.

The lattice measurements discussed  so far are somewhat contradictory.
On one hand, they suggest that the bulk thermodynamical
quantities such as the energy density, the pressure, the entopy, the
susceptibilities, $etc.$ when measured at high temperature are consistent with
the black body limit. On the other hand, the space-like correlators whether
gluonic or hadronic, show strong evidence of correlations and thus an a priori
non-black body limit. How come ?

Before answering this question, let me emphasize the following : all the
lattice calculations performed todate reflect on space-like physics. They have
no bearing on time-like physics. The latter is what matters for
rate calculations at high temperature.

\vskip 1cm
{\bf 3. Hot $O(N)$}
\vskip 3mm

I will now proceed to outline the strategy that I will follow to address the
issues raised by the lattice results. First, I will focus on space-like
physics, second I will work at $T=\infty$ and then expand in $1/T$. The motto
for this approach is dimensional reduction (DR). To illustrate the approach,
I will first discuss it for the $O(N)$ sigma model in two dimensions.

Consider the $O(N)$ model at finite temperature. The field
$\vec \phi$ consists of $N$ bosonic components constrained by
$\phi\cdot\phi=1/g^2$. The Lagrangian density is

\be
{\cal L}= \frac 12 (\partial_{\mu}\vec\phi )\cdot (\partial_{\mu}\vec\phi )
\label{on}
\ee
The free energy of the system reduces to the free energy of $N-1$ free
bosons, to order $g^0$, ${\cal F}/V_1 \sim (N-1)\,{\pi}T^2/6$.
The correlation function can be worked out in powers of $g$. To leading
order and large distances

\be
<\vec{\phi} (x)\cdot\vec{\phi} (0) > = \frac 1{g^2}
\left( 1-\frac 12 g^2 (N-1) T|x| + ...\right) \sim
\frac 1{g^2} e^{-\frac 12 g^2 (N-1) T|x|}
\label{cor}
\ee
These results illustrate rather well the points to be emphasized below.
Indeed, while the free energy reflects on a free bosonic system
to leading order, the correlation function has an exponential fall off
with a correlation length of the order of $1/(N-1)g^2T$. This correlation
length is due to the long wavelength modes in the $O(N)$ model which are
sensitive to the curvature of the $S^{N}$ manifold. Indeed, the Euclidean
partition function reads

\be
Z(T) = \int d\vec\phi\,\,\delta (\vec\phi^2-\frac 1{g^2})\,\,
e^{-\int_0^{1/T} d\tau\int dx\,\frac 12 (\partial_{\mu}\vec\phi )^2}
     = \int d\lambda ({\rm det} (-\nabla^2 +\lambda ) )^{N/2}
       e^{-\frac 1{g^2}\int \lambda\, dx}
\label{part}
\ee
The second equality follows from the integration over the $\vec{\phi}$
field after introducing a Lagrange multiplier $\lambda$.
In the large $N$ limit, the saddle point approximation to (\ref{part})
gives

\be
N\frac{T}{2\pi}\sum_{-\infty}^{+\infty}\int_0^{\Lambda}
\frac{dk}{k^2 + (2\pi n T)^2 +\lambda }=\frac 1{g^2}
\label{gap}
\ee
which can be rearranged to give

\be
\frac{NT}{2\sqrt{\lambda}} = \frac 1{g^2(1+\frac N{2\pi}g^2{\rm
ln}(\frac{T}{\Lambda}))}=\frac 1{g^2(T)}
\label{leng}
\ee
the right hand side of (\ref{leng}) is the $n=0$ contribution in (\ref{gap}).
The nonzero modes in (\ref{gap}) renormalize the coupling constant
$g^2\rightarrow g^2(T)$.

To enhance the physics of the nonzero modes it is more efficient to take the
$T=\infty$ limit first, in which case the high temperature problem for the
$O(N)$ model reduces to a quantum mechanics problem
(field theory in $0+1$ dimensions). Indeed,
on the strip $[0,1/T]\times R$, the bosonic fields are periodic,

\be
\vec\phi (\tau, x) = \vec\phi_0 (x) +\sum_{n\neq 0} e^{-i2\pi nT\tau}
\vec\phi_n (x)
\label{fou}
\ee
If we choose to redefine $\vec{\phi}_0 (x)\rightarrow \vec{x} (t) T$,
then at high temperature only the zero modes contribute to $Z(T)$,

\be
Z(T\rightarrow\infty )= \int d\vec x \,\,\delta (\vec{x}^2 -\frac
1{g^2T^2})\,\,
e^{-\int_0^{1/T}dt\, \frac T2 {\dot{x}}^2}
\label{lar}
\ee
This is the partition function of a quantum mechanical particle of mass $T$
on a sphere of
radius $R=1/gT$. The Hamiltonian is $H=\vec{L}^2/2TR^2$ where
$\vec{L}$ is the angular
momentum in N-dimensions. The eigenstates of the Hamiltonian are hyperspherical
harmonics and the spectrum is given by $E_n =n(n+N-2)/2TR^2$.
There is a gap in the spectrum. The constant modes contribute zero to the free
energy to leading order, but dominate the correlation function

\be
T^2\,<0|\vec{x}(t)\cdot\vec{x}(0)|0>\sim e^{-E_1 t}
\sim e^{-\frac 12(N-1)g^2T\,t}
\ee

\vskip 1cm
{\bf Hot { QCD}$_3$}
\vskip 3mm

I now proceed to discuss the high temperature limit of three dimensional
QCD. The dimensional reduction scheme in QCD$_3$ is exact, as opposed to
QCD$_4$ which has shortcomings beyond one-loop. Many of the results derived
in three dimensions are emmenable to four dimensions. Also the three
dimensional theory has the advantage of being easier to track down
numerically.

Note that the pure Yang-Mills version of QCD$_3$ is also
$Z_N$ symmetric. If the deconfinement phase transition is indeed related
to the spontaneous breaking of $Z_N$ at high temperature, then the three
dimensional theory should display also a phase transition that could be
mapped onto the $Z_N$ Potts model. In this case for both $N=2,3$ the
transition could be second order. In three dimensions parity could be
spontaneously broken. If we were to assume that a condensate forms at low
temperature and disappears at high temperature, then there is a possibility
of a $Z_2$ transition related to parity restoration. If confirmed, this
transition is of the Ising type in two dimensions.

At high temperature QCD$_3$ dimensionally reduces to QCD$_2$ plus a massive
Higgs. To see this consider three dimensional QCD on the cylinder
$[0,1/T]\times R^2$

\be
{\cal L}=\frac 14 F_{\mu\nu}^2 + \psi^+ (-i\rlap/\partial + g_3\, A )\psi
\label{l3}
\ee
In three dimensions QCD is superrenormalizable. $g_3$ is a dimensionful
coupling constant. The gauge fields obey periodic boundary conditions
modulo periodic gauge transformations,  and the fermions antiperiodic
boundary conditions. Explicitly

\be
&& A_{\mu}(\tau, x) = A_{\mu, 0} (x) + \sum_{n\neq 0}e^{i2\pi n T \,\tau}
A_{\mu,m}(x)\nonumber\\ &&
\psi (\tau, x) =\sum_{\pm} e^{\pm i\pi T\tau}\psi_{\pm} +
\sum_{n\neq 0,-1} e^{i(2n+1)\pi T\, \tau}
\label{fourier}
\ee
At high temperature the theory truncates to $R^2$ with massless magnetic
gluons $A_i$, massive electric gluons $A_0$ and "heavy fermions" with mass
$m=(2n+1)\pi T$ [9],

\be
{\cal L}= && \frac 14 F_{ij}^2 +
\frac 12 |(\partial_i -g_3\sqrt{T} A_i )\phi |^2
+\frac 12 m_H^2 \phi^2 + V_H (\phi )\nonumber\\ &&
+\psi^+ (-i\rlap/\partial + g_3\sqrt{T} A +\gamma^3 m +\gamma^3 g_3\sqrt{T}\phi
)\psi
\label{DR}
\ee
We have rescaled the fields to their canonical dimensions in two dimensions,
$A_3\rightarrow\sqrt{T} \phi$, $A_i\rightarrow \sqrt{T} A_i$ and
$\psi\rightarrow \sqrt{T}\psi$. The screening mass $m_H$ is infrared sensitive
in perturbation theory. Its self consistent determination will be discussed
below.

The effective potential $V_H$ for the Higgs is induced by integrating over the
magnetic gluons as shown in Fig.7. Dimensional arguments yield [9]

\be
V_H (\phi ) = \sum_{n\leq 3} C_n \frac{(g_3\sqrt{T} )^n}{T^{n-2}} \phi^n
\label{higgs}
\ee
At high temperature and for a fixed screening mass (to be specified below),
the Higgs potential is subleading in $1/T$. Thus to leading order the Higgs
field is massive and interacts solely with the magnetic gluons and the
"heavy" fermions. Note that the fermions carry an energy $m$. However through
a pertinent $\gamma_3$ rotation of the spinors, this energy can be turned
to a mass [9]. This will be assumed throughout.

In the high temperature limit, the Polyakov line simplifies to

\be
P(x) = {\rm Tr{\bf P}}e^{ig_3\int_0^{1/T} A_0 (\tau, x) \,d\tau}
     \sim {\rm Tr} e^{i\frac{g_3}{\sqrt T}\phi (x)}
\ee
To leading order in $1/T$ the connected part of the Polyakov-anti-Polyakov
correlator is given by

\be
<P(x)\,P^+(0)>_c \sim \frac{(g_3\sqrt{T})^4}{16T^4} <\phi^2 (x)\,\phi^2 (0)>_c
\label{popo}
\ee
This is the correlation function for two electric gluons. The correlator
is dominated by the diagrams shown in Fig.8. The asymptotic form
of the correlator in Euclidean space is just the tail of the closest
singularity to zero in Minkowski space. This singularity corresponds
to a bound state in $1+1$ dimension. The bound state equation is given by

\be
-\frac 1{m_H}\psi" (x) +\frac 12 g_3^2NT |x| \psi (x) = E \psi (x)
\label{shro}
\ee
in the regime where $m_H/T << 1$ (nonrelativistic limit).
The linear potential reflects on the Coulomb potential in $1+1$
dimension. The solutions to (\ref{shro}) are Airy functions. The spectrum
follows from the zeros of the derivative of the Airy function
$Ai' (- E_n ) =0$ (after proper rescaling). The lowest state

\be
E_0\sim \left(\frac{g_3^2T}{\sqrt{m_H}}\right)^{2/3}
\ee
determines the slope of the correlation function (\ref{popo}). Indeed,
at high temperature we expect

\be
<P(x)\,P^+(0)> \sim K_0 ((2m_H+E_0) |x|)\sim
\frac{e^{- (2m_H + E_0) |x|}}{({(2m_H+E_0)|x|})^{1/2}}
\label{one}
\ee
which is to be contrasted with the free screening correlator (free bubble
of Fig.8)

\be
<P(x)\,P^+(0)> \sim K^2_0 ( m_H |x|)\sim
\frac{e^{- 2m_H |x|}}{ m_H|x|}
\label{two}
\ee
Note that at high temperature the difference between (\ref{one}) and
(\ref{two}) is in the preexponent and not the exponent,
since $E_0/m_H\rightarrow 0$. For QCD in four dimensions $K_0\rightarrow K_1$.

In $QCD_3$, the screening mass $m_H\sim g_3^2 T$ is infrared sensitive
and gauge dependent in perturbation theory as indicated by the graph of
Fig.9. A self-consistent and gauge independent analysis of the screening
mass can be performed from the loop-expansion of the Polyakov-Polyakov
correlator at high temperature. The two-loop contribution to
(\ref{popo}) is shown in Fig.10. The cross refers to the $-m_H^2$
insertion in the self-consistent (Hartree) definition of the electric mass,
where the electric propagators carry a mass $m_H^2$. All the infrared
and ultraviolet divergences cancel to two-loop. The finite contributions
from diagram 10a  and 10b  are scale ($\mu $) dependent :
$g_3^2 NT (a+b{\rm ln}({m_H}/{\mu}))$. The finite contribution of diagram
10c is also scale dependent
$g_3^2 NT (\overline{a} + \frac 1{2\pi}{\rm ln}({m_H}/{\mu}))$.
If we choose $\mu = m_H$ then the contribution from 10c can be made
logarithmically large compared to the contributions from 10a and 10b
in the large temperature limit. This large contribution can be reabsorbed
in a self-consistent definition of the electric mass in a Hartree-type
approximation,

\be
-m_H^2 + g_3^2 N T\frac 1{2\pi}{\rm ln}\left( \frac{T}{m_H}\right) = 0
\label{hartree}
\ee
This result was first obtained by D'Hoker [10].

The fact that magnetic loops still obey an area law has a simple explanation in
high temperature QCD in three dimensions. Indded, to leading order the static
part of the magnetic gluon correlator in static-axial ($A_2=0$)
gauge follows from the diagrams of Fig.11 ,

\be
\Delta_{ij} \sim \delta_{i1}\delta_{j1} \delta (x_1)
\left( a |x_2| + \frac c{\sqrt{Mx_2}}e^{-M |x|}\right)
\label{pro}
\ee
where $a= g_3^2NT/24\pi m_H^2$, $b=g_3^2NT/240\pi m_H^4$, $c$ an arbitrary
coefficient and $M^2=(1-a)/b$. In (\ref{pro}) only the leading temperature
dependent part has been retained. The contribution of (\ref{pro}) to the
space-like Wilson loop is due to the unscreened part of the magnetic
gluon propagator. The result is $W_{xy} \sim e^{-\sigma_{xy} A}$,
where $A$ is the area in the xy-plane and $\sigma\sim a$ the $temperature$
dependent string tension to leading order. This result is gauge invariant.
On the other hand, note that the magnetic-magnetic correlator is
exponentially screened. At large distances, it is dominated by the second term
of (\ref{pro}) in the static-axial gauge

\be
<B(x)\cdot B(0)> \sim \frac{ e^{-M|x|}}{\sqrt{Mx_2}}
\ee
The linear term drops by taking derivatives. This result is gauge dependent.
Note that an area law behavior of the space-like Wilson loops is still
consistent with an exponential fall-off in the correlation function of
magnetic fields. I expect this to extend to four dimensions as well.

The static hadronic correlators can be analysed in the dimensionally
reduced theory, by treating the fermions as infinitely heavy
(non-relativistically). To leading order in $1/T$

\be
C_{\alpha , \alpha} (x) =\frac 1{\beta^2}
\langle \int_0^{\beta} d\tau\psi^+\Gamma^{\alpha}\psi (\tau, x)
 \int_0^{\beta} d\tau'\psi^+\Gamma^{\alpha}\psi (\tau', 0)\rangle
\sim
\langle \psi^+\Gamma^{\alpha}\psi (x) \,
\psi^+\Gamma^{\alpha}\psi (0)\rangle
\label{dr2}
\ee
The fermions of the second part of (\ref{dr2}) are restricted to the lowest
Matsubara modes and carry a mass $m=\pi T$ after a $\gamma_3$-rotation.
The dominant contribution to the hadronic correlator (\ref{dr2}) follows from
the diagram of Fig.12. The Higgs and fermion insertions are subleading in the
temperature. The large $x$-asymptotics of (\ref{dr2}) is again given by
the closest singularity to 0 in Minkowski space. In Minkowski space the quark
and antiquark are massive (nonrelativistic) and interact via a Coulomb
potential in $1+1$ dimensions. The bound state equation is

\be
-\frac 1{m}\psi'' +\frac {e^2}2 |x|\psi = E_{\alpha} \psi
\label{shro2}
\ee
where $e^2=C_Fg_3^2 T$ and $C_F$ the value of the Casimir in the
fundamental representation. The screening lengths are
$m_{\alpha}=2m+E_{\alpha}$, and characterize
$C_{\alpha\alpha}(x\rightarrow \infty )= e^{-m_{\alpha}|x|}$.
For QCD$_3$ [9],

\be
m_{\alpha} = 2\pi T +\frac 32 \left(\frac{C_F^2g_3^4T}{4\pi}\right)^{1/3}
\ee
The wavefunctions are Airy functions with a size of the order of
$1/(g_3T)^{2/3}$. These wavefunctions are directly related to the
transverse correlations. For four dimensional QCD, the screening
lengths can also be estimated [9]. The reduced wavefunctions directly
relate to the correlations observed on the lattice as described above.
The present discussion provides a simple explanation to the fact that
while the screening lengths asymptote "free" quark values, they are in fact
reflecting on strong correlations space-like. The correlations are subleading
in the screening masses.

The size of the fermionic distribution around a heavy source has a simple
understanding in the dimensionally reduced theory. Indeed, at high temperature

\be
<P(0)\overline{\psi}\psi (x)>\sim -\frac{g_3^2}T <\phi^2 (0)
\overline{\psi}\psi (x)>
\label{cloud}
\ee
At large temperature the Polyakov line is a source for the Higgs field.
The correlation function (\ref{cloud}) measures the fermionic distribution in
the Higgs cloud. The range of the correlation is of the order of the inverse
of twice the electric length $1/2m_H$.

While the above considerations have provided a large body of evidence for why
the space-like physics at high temperature reflects on correlations, it remains
puzzling why the bulk properties of the high temperature QCD phase are
consistent to some extent with the black-body description. I will show now that
this is in fact a consequence of the fact that the bulk quantities reflect on
$all$ distance scales. For that consider the isoscalar fermionic susceptibilty

\be
\chi (T) = \int_0^{1/T}d\tau\int d^2x
\langle \psi^+\gamma^3\psi (\tau, x )\psi^+\gamma^3\psi (0,0)\rangle
\sim T\int d^2x \langle \psi^+\sigma^3 \psi (x) \psi^+ \sigma^3\psi (0)\rangle
\ee
$All$ fermionic modes contribute to $\chi (T)$ to leading order. Indeed [11],

\be
\chi (T) =\frac {N_cT}{\pi}\sum_{H=-\infty}^{+\infty}\sum_{n=0}^{+\infty}
{\rm sin}^2(\frac{n\pi}2)\frac{G_{nH}^2}{\mu_{nH}^2}
\label{sus}
\ee
The $\mu$'s are the invariant masses for bound quark-antiquark in
$1+1$ dimensions with masses $m_F = (2F+1)\pi T$, and $G_{nH}$ the
form factors following from the t'Hooft equation. Asymptotically
$\mu_{nH}^2\rightarrow \pi\sqrt{2\pi n\sigma}/2\sigma$, $\sigma = g_3^2NT$
and $G_{nH}\rightarrow \pi/\sqrt{2}$. The density of states per unit energy
in (\ref{sus}) grows linearly with the energy
$dn/dE\sim \sqrt{E^2-4m_F^2}/\pi\sigma$. With this in mind, the sums in
(\ref{sus}) can be done exactly, leading to [11]

\be
\chi (T)- \chi (0) \sim \frac{N}{\pi}\int
\frac{dz}{e^{z/T}+1} = N_c\left(\frac{{\rm ln}2}{\pi}\right) T
\ee
which is the free gas result to leading order. I believe this result
to extend to the bulk energy density, pressure and entropy.

\vskip 1cm
{\bf 5. Conclusions}
\vskip 3mm

A global understanding of the lattice calculations can be achieved within the
context of dimensional reduction at very high temperature. Eventhough the
analytical calculations were performed for high temperature QCD$_3$, they can
be generalized in a straightforward way to QCD$_4$. However, the arguments are
not rigorous in the latter.

We have seen that the high temperature QCD phase is characterized by strong
correlations space-like. These correlations are dominant in all correlators.
Bulk quantities, however, are sensitive to all correlated modes. The latters
sum up to the expected black body limits, a result familiar from asymptotic
freedom when hadronic wavefunctions are probed in deep Euclidean region.

What does this imply for the high temperature phase time-like ? I really
do not know. However, since the correlators are not free space-like
why should they be time-like ?

\vglue 1cm
{\bf \noindent  Acknowledgements } \\
\noi
Many of the ideas discussed here were developed in collaboration
with Hans Hansson. I thank the organisers of the Zakopane school for a very
pleasant meeting. This work was supported in part by the
US Department of Energy under Grant No. DE-FG-88ER40388 and KBN grant No
PB 2675/2.

\vglue 1cm
{\bf \noindent  References}
\vskip .5cm

\noindent [1] B. Petersson, Nucl. Phys. (Supp) {\bf 30B} (1993) 66

\noindent [2] J.B. Kogut, Phys. Rev. Lett. {\bf 56} (1986) 2557.

\noindent [3] F. Karsch, Nucl. Phys. (Supp) {\bf 9B} (1989) 357.

\noindent [4] S. Gottlieb, $et\, al.$, Phys. Rev. Lett. {\bf 59}  (1987) 2247.

\noindent [5] B. Peterson, private communication.

\noindent [6] R.V. Gavai $et \, al.$ Phys. Lett. {\bf 241B} (1990) 567.

\noindent [7] C. Bernard $et \, al.$ Phys. Rev. Lett. {\bf 68} (1992) 2125.

\noindent [8] C. DeTar, Nucl. Phys. (Supp) {\bf 30B} (1993) 66.

\noindent [9] T.H. Hansson and I. Zahed, Nucl. Phys. {\bf 374B} (1992) 177.

\noindent [10] E. DeHoker, Nucl. Phys. {\bf 201B} (1982) 401.

\noindent [11] M. Prakash and I. Zahed, Phys. Rev. Lett. {\bf 69} (1992) 3282.

\newpage
{\bf Figure Captions}
\vskip 9cm
\noindent Fig. 1 : Chiral order parameter as a function of $\beta =6/g^2$
                   for 4 flavours and quark masses $m_qa = .025$
                   (triangles), $.01$ (circles) and 0 (squares) [1].

\vskip 9cm
\noindent Fig. 2 : Energy density and pressure normalized to the
                   black body limit on a finite lattice versus
                   $T/T_c$ [3]. The dashed horizontal lines refer
                   to the perturbative corrections to the black body
                   limit.
\newpage
\vspace*{ 9 cm}
\noindent Fig. 3 : Singlet ($\chi_S$) and triplet ($\chi_{NS}$)
                   as a function of $\beta =6/g^2$ for two flavours [4].

\newpage
\vspace* {9 cm}
\noindent Fig. 4 : (a) Screening masses $\mu/T =m_H/T$ as a function of $\beta$
for
                   four flavours and $m_qa=.01$ [6]. $\beta =5.15$ is the
                   transition point. (b) the same as (a) but for the $\pi$
                   and $\sigma$ propagators in the spatial directions.

\vskip 9cm
\noindent Fig. 5 : the pion (PS) and $\rho$ (VT) "wavefunctions" at $T=0$
                   and $T=1.5$ [7].

\newpage
\vspace* {9 cm}
\noindent Fig. 6 : Quark number density induced by a fixed quark at the origin
                   at three temperatures [8].
\newpage
\vspace* {5 cm}
\noindent Fig. 7 : Induced Higgs potential $V_H$ by dimensional reduction.
                   The wavy lines refer to the magnetic gluons.
\vskip 7cm
\noindent Fig. 8 : Leading contribution to the Polyakov-anti-polyakov
                   correlator (connected) in the dimensionally reduced
                   theory.
\newpage
\vspace* {5cm}
\noindent Fig. 9 : One-loop contribution to the screening mass.

\vskip 9cm
\noindent Fig. 10 : Two-loop contributions to the Polyakov-anti-Polyakov
                    correlator (connected).
\newpage
\vspace* {5 cm}
\noindent Fig. 11 : Magnetic polarisation in the dimensionally reduced theory,
                    to leading order.

\vskip 9cm
\noindent Fig. 12 : Hadronic correlators in the dimensionally reduced theory.

\end{document}